%% file: 0_main.tex
\documentclass{article}
\usepackage{spconf}
\usepackage{multirow}

\usepackage{graphicx} 
\usepackage{enumitem}
\setlist{nosep, leftmargin=14pt}
\usepackage{paralist}
\usepackage{xcolor}
\usepackage{tabularx}
\usepackage{booktabs}

\usepackage{amsmath}
\usepackage{amssymb}
\usepackage{mwe} 
\usepackage{bbm}
\usepackage{ifthen}
\usepackage{adjustbox}
\usepackage{xkeyval}
\usepackage{tikz}
\usepackage{calc}
\usepackage{comment}

\usepackage[hidelinks]{hyperref}
\usepackage{xspace}
\usepackage{bm} 
\usepackage{colortbl}
\usepackage[normalem]{ulem}
\useunder{\uline}{\ul}{}
\usepackage{lipsum}

\newcommand\blfootnote[1]{%
  \begingroup
  \renewcommand\thefootnote{}\footnote{#1}%
  \addtocounter{footnote}{-1}%
  \endgroup
}

\usepackage[capitalize]{cleveref}
\crefname{section}{Sec.}{Secs.}
\Crefname{section}{Section}{Sections}
\Crefname{table}{Table}{Tables}
\crefname{table}{Tab.}{Tabs.}

\def\etal{\emph{et al. }}
\def\eg{\emph{e.g. }}

\newcommand{\todo}[1]{{\textcolor{red}{#1}}}

\newcommand{\convabb}{SRE\xspace}
\newcommand{\convname}{SRE-Conv\xspace}

\title{Improved Vessel Segmentation with Symmetric \\Rotation-Equivariant U-Net}
\name{
Jiazhen~Zhang$^{1}$, 
Yuexi~Du$^{1}$, 
Nicha~C.~Dvornek$^{1,2}$,
John~A.~Onofrey$^{1,2,3}$
}
\address{
    Departments of 
    $^1$Biomedical Engineering, 
    $^2$Radiology \& Biomedical Imaging, \\
    $^3$Urology, 
    Yale University, New Haven, CT, USA \\
}
\begin{document}
%
\maketitle
\begin{abstract}
Automated segmentation plays a pivotal role in medical image analysis and computer-assisted interventions.
Despite the promising performance of existing methods based on convolutional neural networks (CNNs), they neglect useful equivariant properties for images, such as rotational and reflection equivariance. 
This limitation can decrease performance and lead to inconsistent predictions, especially in applications like vessel segmentation where explicit orientation is absent. 
While existing equivariant learning approaches attempt to mitigate these issues, they substantially increase learning cost, model size, or both.
To overcome these challenges, we propose a novel application of an efficient symmetric rotation-equivariant (\convabb) convolutional (\convname) kernel implementation to the U-Net architecture, to learn rotation- and reflection-equivariant features, while also reducing the model size dramatically. 
We validate the effectiveness of our method through improved segmentation performance on retina vessel fundus imaging.
Our proposed \convabb U-Net not only significantly surpasses standard U-Net in handling rotated images, but also outperforms existing equivariant learning methods and does so with a reduced number of trainable parameters and smaller memory cost.
The code is available on \url{https://github.com/OnofreyLab/sre_conv_segm_isbi2025}.

\end{abstract}
\blfootnote{\textcopyright~2025 IEEE. Personal use of this material is permitted. Permission from IEEE must be obtained for all other uses, in any current or future media, including reprinting/republishing this material for advertising or promotional purposes, creating new collective works, for resale or redistribution to servers or lists, or reuse of any copyrighted component of this work in other works.}


\input{1_introduction.tex}
\input{2_method.tex}
\input{3_results.tex}
\input{4_discussion.tex}

\input{5_acknowledgements.tex}


\bibliographystyle{IEEEbib}
\bibliography{strings,refs}

\end{document}

%% file: 1_introduction.tex
\section{Introduction}
\label{sec:intro}

\begin{figure}[t]
    \centering
    \includegraphics[width=0.95\columnwidth]{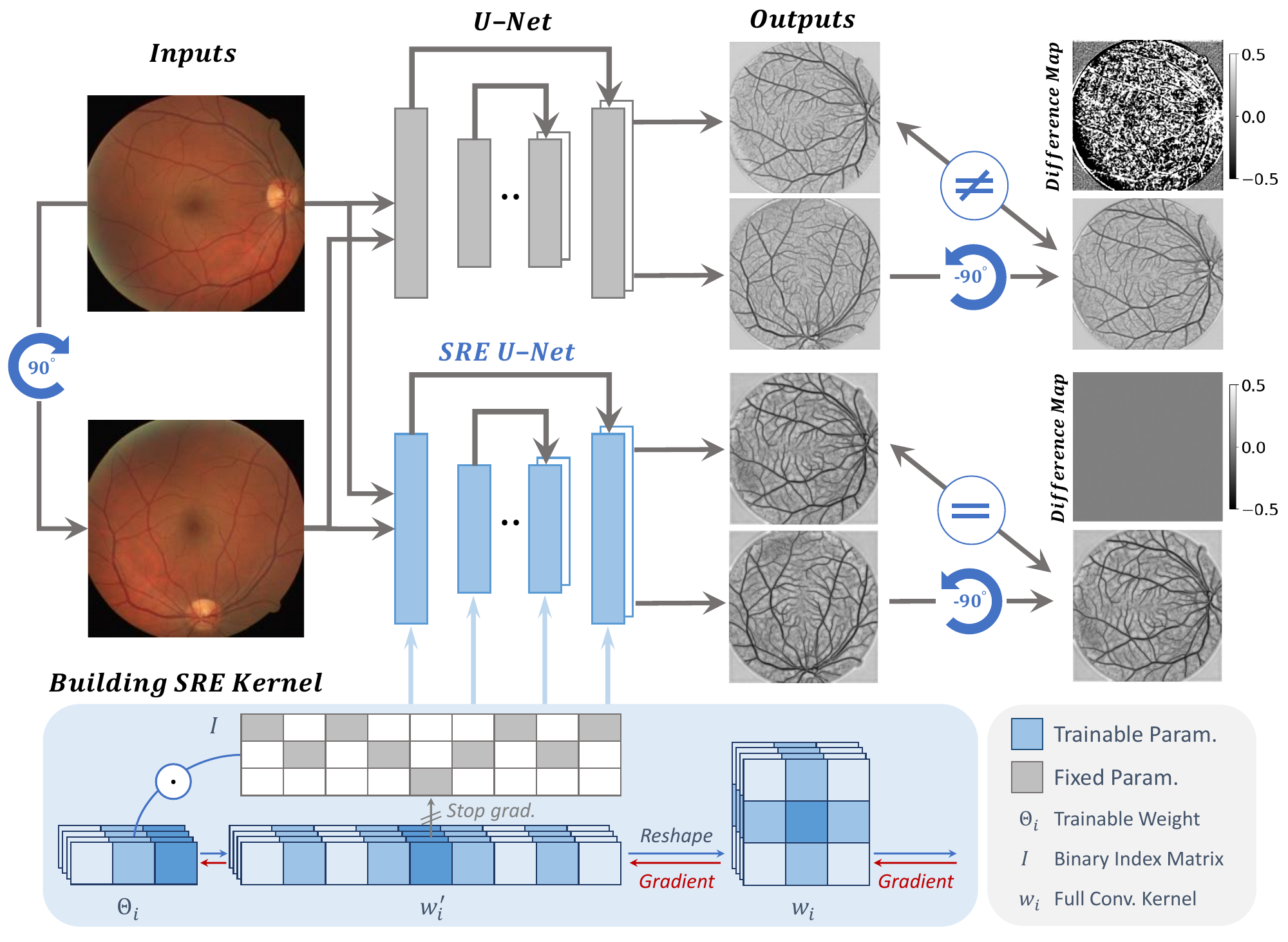}
    \vspace{-1.0\baselineskip}
    \caption{
        \textbf{Symmetric Rotation-Equivariant (SRE) U-Net.}
        Compared to standard U-Net, \convabb U-Net utilizes a parameter-efficient symmetric kernel to enable rotational equivariance. Difference image maps between the output of the original and rotated (90{\textdegree}) images illustrate \convabb U-Net's equivariant property by maintaining consistent feature response after rotation on a 2D fundus image.
    }
    \vspace{-1.0\baselineskip}
    \label{fig:figure1}
\end{figure}

Segmentation is a fundamental task in medical imaging analysis that involves identifying and delineating regions of interest, such as organs, lesions, and tissues.
Accurate segmentation is essential for many clinical applications, including disease diagnosis, treatment planning, and monitoring of disease progression.
Deep convolutional neural networks (CNNs)~\cite{ronneberger2015u,ji2021learning,isensee2021nnu, zhang2022atlas} have shown great promise in segmenting medical images due to their ability to learn intricate image features and deliver accurate segmentation results across a diverse range of tasks.
Standard convolutional operations exhibit translational equivariance, which permits efficient detection of similar features across various input positions.
The capability of CNNs, however, degrades when images are rotated or flipped, as standard convolutional kernels are not equivariant to rotations and reflections.
An illustration (\cref{fig:figure1}) of this limitation is observed when a standard CNN processes an image rotated by 90{\textdegree}. 
Here, the output feature maps differ substantially from those produced by the original inputs when rotated back to the original orientation. 

Data augmentation strategies, such as random rigid transformations and reflection, can boost segmentation model performance by effectively increasing the number of training data samples. 
However, this approach does not enforce the rotation or reflection equivariance of the feature maps.
Alternatively, the convolution kernels could be actively rotated at each layer in the network to encode feature maps for different orientations~\cite{dieleman2016exploiting,chidester2019rotation,cohen2016group} instead of implicitly learning equivariance by rotating input images. Group operation-based rotation equivariant convolution has previously been extended to semantic segmentation~\cite{linmans2018sample,winkens2018improved,pang2022beyond,bernander2022rotation}. 
However, these methods incur heavy computational costs, in which computation of the rotation operations dramatically increases memory requirements by repeating rotated feature outputs at each layer. 
Other approaches seek to achieve rotation-equivariance through spherical harmonics~\cite{worrall2017harmonic} and steerable kernels~\cite{weiler2018learning,weiler2019general,cesa2022a} that encode pre-defined rotation angles. 
However, these methods also have high computational costs and fail to generalize to undefined angles. 
Rotation equivariance can also be achieved by using symmetric convolution kernels~\cite{dudar2019use,fuhl2021rotated}.
By explicitly setting the convolution kernel to have a centrally symmetric form, these symmetric rotation-equivariant (\convabb) convolution (\convname) kernels~\cite{du2025sreconv} can capture rotation-equivariant features (\cref{fig:figure1}).

In this study, we apply \convname to perform efficient, rotation equivariant semantic segmentation.
We develop the \convabb U-Net that benefits from the \textit{plug-and-play} nature of \convname; 
we perform experiments on the public DRIVE retina vessel segmentation dataset~\cite{staal2004ridge}, which consists of rotation-equivariant structures; 
and we provide extensive benchmarking against standard U-Net~\cite{ronneberger2015u} segmentation, equivariant learning methods~\cite{pang2022beyond,weiler2019general}, and state-of-the-art (SoTA) retinal vessel segmentation~\cite{liu2022full}, to assess the efficacy of integrating our equivariant kernels into a semantic segmentation model. 
Quantitative and qualitative results demonstrate that our equivariant semantic segmentation approach accurately segments vessels across rotated image inputs and uses fewer parameters compared to other methods.

%% file: 2_method.tex
\section{Methods}
\label{sec:methods}

\subsection{Symmetric Rotation-Equivariant (\convabb) Convolution}
\label{sec:methods:sre_kernels}

CNNs are equivariant with respect to translation. 
This means that translating the input to a convolutional layer will result in translating the output by the same amount. 
To achieve rotational equivariance, we apply \convname kernels~\cite{du2025sreconv}, which parameterize the kernels to be centrally symmetric. 
\convabb kernels contain many redundant values and are implemented in a parameter-efficient way  (\cref{fig:figure1} shows a $k\times k$ 2D convolutional kernel for illustration).
To implement \convname kernel $w_{i}$ with as few parameters as possible, the method splits the convolutional kernel into $b=\left\lfloor k/2 \right\rfloor+2$ discrete bands, corresponding to $b$ individual trainable parameters, denoted as a matrix $\Theta_{i} \in \mathbb{R}^{ [C,b] }$ where $C$ is the number of feature channels. 
A binary index matrix $I \in \mathbb{R}^{ [b, k^{2}] }$ indicates the bands to which each trainable parameter corresponds.
In this way, the final \convname kernel is given by:
\begin{equation}
    \begin{array}{l}
w_{i}=\psi (w_{i}^{'})=\psi (\Theta_{i} \cdot I)
    \end{array}
\end{equation}
where $\psi(\cdot)$ denotes the reshaping operation and  $\cdot$ denotes matrix multiplication.
By turning convolution into matrix multiplication in the implementation during training, we achieve lower complexity and less number of floating operations (FLOPs) compared to regular convolution. Therefore, \convname is much more efficient with less computational cost than regular convolution. 

\subsection{Network Architecture}
\label{sec:methods:network}

We implement our \convabb semantic segmentation network using the U-Net~\cite{ronneberger2015u} architecture as our backbone.
We replace all standard convolution kernels in both the encoder and decoder with  \convname kernels.
In the encoder, we use max pooling operations to downsample feature maps. We implement two downsampling layers to further compress model size.
In the decoder, in contrast with a standard U-Net that uses trainable up-convolution to upsample the low-resolution feature maps, we upsample feature maps with linear interpolation in order to maintain the rotation equivariance of the feature maps. 
This ensures the strict rotation equivariant property of the network since trainable up-convolution layers with a stride larger than 1 will break the rotation equivariant property by convolving at different positions after rotation.

%% file: 3_results.tex
\section{Experiments and Results}
\label{sec:results}

\subsection{Experimental Setup}
\label{sec:results:experimentalsetup}

\paragraph*{Dataset:}
\label{sec:results:datasets}
We evaluate using the public retina vessel DRIVE dataset~\cite{staal2004ridge}, which consists of 40 2D RGB fundus images with paired binary vessel segmentation labels.
We partition the dataset into equal halves for training and testing.

\input{3_Tab2}

\paragraph*{Baselines Comparison:} 
We choose the standard \textbf{U-Net}~\cite{ronneberger2015u} with the same architecture as our \textbf{\convabb U-Net} for our primary baseline. 
To evaluate performance with respect to equivariant learning approaches, we compare the following baseline methods: \textbf{Group U-Net}~\cite{pang2022beyond}, and \textbf{ES U-Net}~\cite{weiler2019general,cesa2022a}. These methods also use the same U-Net architecture. 
Furthermore, we compare to \textbf{FR U-Net}~\cite{liu2022full}, the SoTA retinal vessel segmentation method on DRIVE, as well as the non-deep learning filter-based \textbf{Frangi}~\cite{frangi1998multiscale} segmentation method.
All deep learning methods use the same hyperparameter settings, data augmentation, and training from scratch. 
Random 90{\textdegree} rotations and flipping are applied in all experiments to augment the dataset and avoid interpolation artifacts.

\paragraph*{Evaluation Metrics:}
We evaluate segmentation performance on the predicted segmentation maps of the test set with the ground-truth labels by assessing \textbf{accuracy}, \textbf{sensitivity}, \textbf{specificity}, intersection over union (\textbf{IoU}), \textbf{Dice} coefficient, and the area under the receiver operating characteristic curve (\textbf{AUC}). 
In addition to evaluating the original test images (0{\textdegree}), we rotate the test set images in 1{\textdegree} increments within the range $\pm5^{\circ}$ to form a rotated test set. 
Rotated images are resampled using nearest-neighbor interpolation to avoid artifacts.
This test set evaluates the model's ability to maintain rotational equivariance under small angle perturbations. 
Such small angle perturbations realistically simulate varying imaging conditions common to fundus imaging, and can be critical in real-world diagnostic applications.

\paragraph*{Experimental Settings:}
\label{sec:results:settings}
We implement our SRE U-net using \convname with sizes $k$=[9,7,5] at each layer (ablation studies in 
\cref{sec:results:quan_eval:ablation}
justify our model architecture).
We train the network in a supervised manner by minimizing the cross-entropy loss between the ground-truth segmentation labels and predicted results. 
Data augmentation of random flipping and 90{\textdegree} rotation is performed to enhance training data variability. 
Patch size is set to 96$^2$ and training uses a batch size of 32.
AdamW optimization is used with an initial learning rate set to 5e-4 and cosine-annealing decay.
Training the network requires 6,000 epochs. 
Our method is implemented with PyTorch and all experiments use an NVIDIA A5000 GPU.

\subsection{Quantitative Evaluation} 
\label{sec:results:quan_eval}

\begin{figure}[t]
    \centering
    \includegraphics[width=1.0\columnwidth]{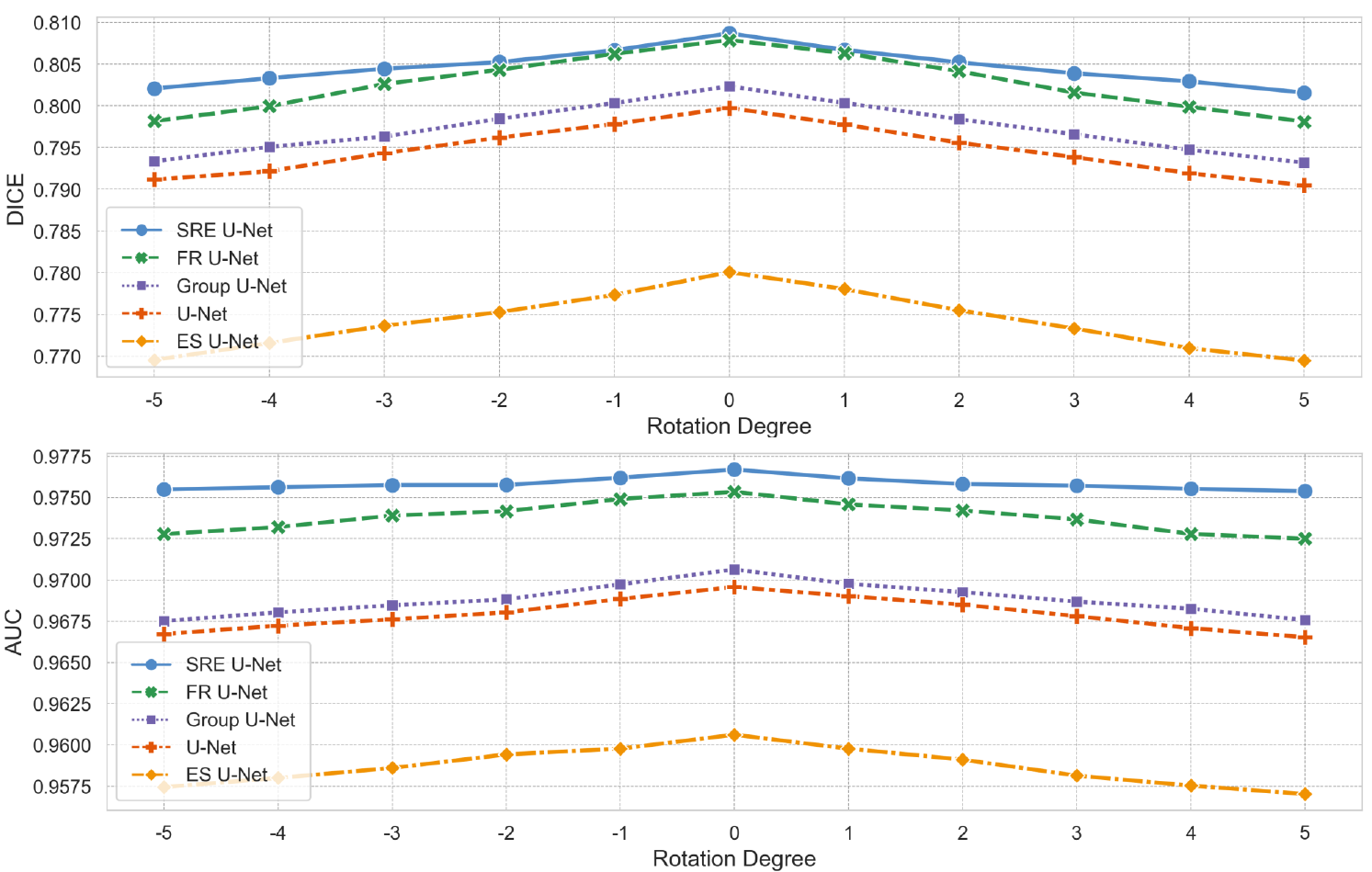}
    \vspace{-2.0\baselineskip}
    \caption{
        \small
        \textbf{Performance Across Rotation Degree.}
        We plot Dice and AUC across each rotation degree to visualize the influence of small rotation on the model's performance.
        }
    \vspace{-1.0\baselineskip}
    \label{fig:figure2}
\end{figure}

\begin{figure*}[t]
    \centering
    \includegraphics[width=0.93\textwidth]{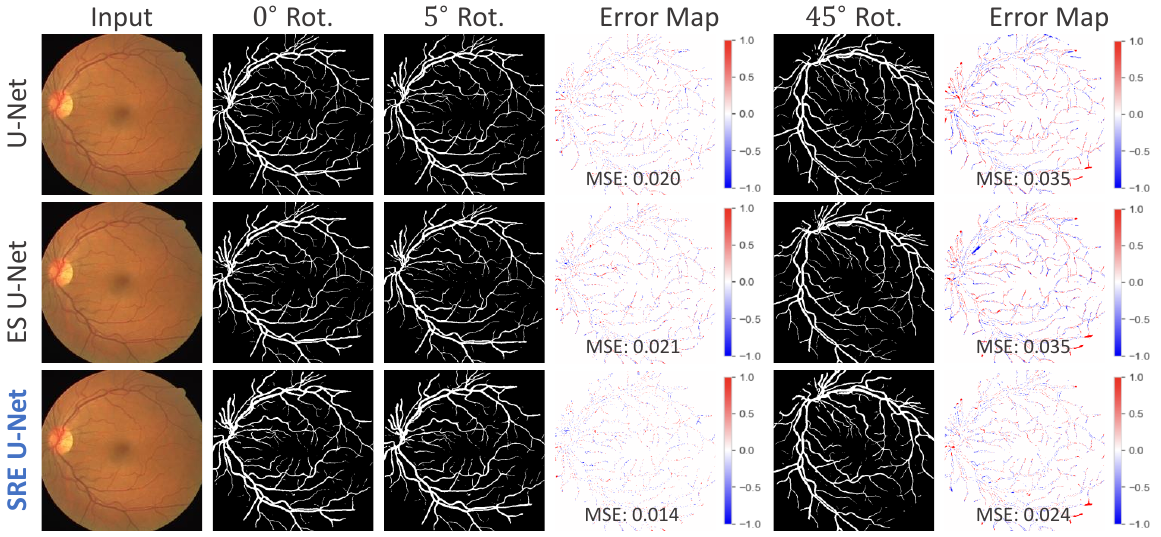}
    \vspace{-1.0\baselineskip}
    \caption{
    \textbf{Qualitative Results.} An example fundus image is rotated and fed into the models and the outputs are rotated back to compare with the output from the original image. Difference maps visualize the difference in the predictions under different rotational conditions and we quantitatively summarize this error using Mean Squared Error (MSE). Our proposed method shows consistent results due to its rotational equivariance property. 
    }
    \vspace{-1.0\baselineskip}
    \label{fig:figure3}
\end{figure*}

\paragraph*{Segmentation Evaluation:} 
\Cref{tab:table2} details model performance on the DRIVE dataset. 
The proposed \convabb U-Net achieves the best performance across most metrics, except Sensitivity, compared to other methods. 
While FR U-Net performs slightly better in Sensitivity, SRE U-Net provides a balanced improvement across all metrics, highlighting its ability to maintain robust segmentation performance even with a smaller number of parameters (0.12M), particularly in handling rotational variations ($\pm5^{\circ}$). 
This suggests our \convabb U-Net can greatly improve both parameter efficiency and prediction consistency, affirming the value of incorporating rotation-equivariant features into CNNs. 
The non-deep learning Frangi filter underperforms all the other methods except for Specificity. 
The Frangi method misses many small vessels, leading to high true negatives (TN) but low false positives (FP).
To assess robustness to small, realistic rotations, we visualize Dice and AUC for each rotation degree (\cref{fig:figure2}). 
These results highlight the stark performance disparity in standard U-Net when test data is rotated.
Our proposed \convabb U-Net effectively maintains its performance with a minimal performance drop.

\subsection{Qualitative Evaluation} 
We visualize the predicted segmentation map for a randomly selected image given example rotations (\cref{fig:figure3}). We compute the difference map after rotating the predicted segmentation map back to the original orientation. 
The proposed \convabb U-Net demonstrates remarkable consistency in performance under both small and large degrees of rotation. 
The difference maps qualitatively highlight the disparity among the three methods, and demonstrate \convabb U-Net's robust rotational equivariance. 
In contrast, the standard U-Net exhibits a marked inability to maintain consistent segmentation predictions as the orientation of the input image varies. 
Notably, this prediction discrepancy becomes increasingly pronounced with greater degrees of image rotation, underscoring a significant limitation in the standard U-Net to handle rotational variations in the data.
The performance of ES U-Net is similar to standard U-Net, indicating the model over-fits to specific pre-defined orientations.

\subsection{Ablation Studies} 
\label{sec:results:quan_eval:ablation}
We investigate the impact of various \convname kernel sizes on model performance (\cref{tab:table1}). 
Notably, increasing the kernel size for each downsampling layer in our model does not always improve the performance.
Larger kernel sizes increase the receptive field, allowing the model to capture more global context. 
However, excessively large kernels can dilute finer details essential for tasks like vessel segmentation, where local structures are crucial. 
This can lead to overfitting, especially with small datasets, as the model captures irrelevant features instead of focusing on critical local patterns. 
In contrast, a mix of small and moderate kernel sizes 
balances global context and local details, leading to improved performance.

\input{3_Tab1}

%% file: 3_Tab2.tex
\begin{table*}[!t]
    \centering
    \caption{\textbf{Segmentation Results.} We report the accuracy, sensitivity, specificity, IoU, Dice, and AUC of the segmentation with each baseline. 
    Metrics are averaged over all rotation angles in the test set.
    The best and second-best results are highlighted in bold and underlined.
    }
    \label{tab:table2}
    \resizebox{\linewidth}{!}
    {
        \begin{tabular}{@{}l c|cc|cc|cc|cc|cc|cc@{}}
        \specialrule{.1em}{.05em}{.05em}
        
        \multicolumn{1}{c}{\multirow{2}{*}{\textbf{Methods}}} & \multicolumn{1}{c|}{\multirow{2}{*}{\textbf{\#Param.}}} & \multicolumn{2}{c|}{\textbf{Accuracy}} & \multicolumn{2}{c|}{\textbf{Sensitivity}} & \multicolumn{2}{c|}{\textbf{Specificity}} & \multicolumn{2}{c|}{\textbf{IoU}} & \multicolumn{2}{c|}{\textbf{Dice}} & \multicolumn{2}{c}{\textbf{AUC}}\\ 
        \cmidrule(l){3-14}
        \multicolumn{1}{c}{} & \multicolumn{1}{c|}{} & ~~$0^{\circ}$~~ & ~~$\pm5^{\circ}$~~ & ~~$0^{\circ}$~~ & ~~$\pm5^{\circ}$~~ & ~~$0^{\circ}$~~ & ~~$\pm5^{\circ}$~~ & ~~$0^{\circ}$~~ & ~~$\pm5^{\circ}$~~ & ~~$0^{\circ}$~~ & ~~$\pm5^{\circ}$~~ & ~~$0^{\circ}$~~ & ~~$\pm5^{\circ}$~~\\  
        
        \specialrule{.08em}{.05em}{.05em} \specialrule{.08em}{.05em}{.05em}

        Frangi ~\cite{frangi1998multiscale}~ & --- & 0.9425 & 0.9422 & 0.6047 & 0.6010 & {\ul 0.9842} & {\ul 0.9842} & 0.5334 & 0.5301 & 0.6943 & 0.6915 & 0.8868 & 0.8856\\ 
        
        U-Net~\cite{ronneberger2015u}~ & {\ul 0.48M} & 0.9583 & 0.9574 & 0.7713 & 0.7641 & 0.9816 & 0.9814 & 0.6671 & 0.6601 & 0.7998 & 0.7947 & 0.9696 & 0.9679\\ 
        
        Group U-Net~\cite{pang2022beyond}~ & 1.92M & 0.9591 & 0.9582 & 0.7678 & 0.7608 & 0.9829 & 0.9828 & 0.6709 & 0.6637 & 0.8023 & 0.7972 & 0.9706 & 0.9688\\
        
        ES U-Net~\cite{weiler2019general,cesa2022a}~ & 0.72M & 0.9554 & 0.9544 & 0.7421 & 0.7335 & 0.9821 & 0.9820 & 0.6427 & 0.6348 & 0.7800 & 0.7740 & 0.9606 & 0.9587\\

        FR U-Net~\cite{liu2022full}~ & 6.97M & {\ul 0.9594} & {\ul 0.9585} & \textbf{0.7922} & \textbf{0.7844} & 0.9803 & 0.9803 & {\ul 0.6789} & {\ul 0.6717} & {\ul 0.8079} & {\ul 0.8027} & {\ul 0.9753} & {\ul 0.9738}\\
        
        \midrule
        
        \convabb U-Net ($k$ list=[9, 7, 5])~ & \textbf{0.12M} & \textbf{0.9600} & \textbf{0.9595} & {\ul 0.7892} & {\ul 0.7837} & \textbf{0.9870} & \textbf{0.9870} & \textbf{0.6795} & \textbf{0.6738} & \textbf{0.8087} & \textbf{0.8046} & \textbf{0.9767} & \textbf{0.9758}\\
        
        \specialrule{.1em}{.05em}{.05em}
        \end{tabular}

    }
    \vspace{-1.0\baselineskip}
    
\end{table*}

%% file: 3_Tab1.tex
\begin{table}[!t]
    \centering
    \caption{\textbf{\convname Ablation Studies.} Model performance using different kernel size configurations, where $k$ Configuration corresponds to the kernel sizes used in each U-net layer. 
    Dice are averaged across rotation angles in the test set. We highlight the best result for each model in bold and the second best with underlining. }
    \label{tab:table1}
    \resizebox{0.75 \linewidth}{!}
    {
        \begin{tabular}{lcc}
            \specialrule{.1em}{.05em}{.05em} 
            
            \textbf{$\bm{k}$ Configuration} & \textbf{\#Param.} & \textbf{Dice ($\bm{\pm5^{\circ}}$})\\
            \midrule \midrule
            {[5, 5, 5]} & 0.09M & 0.7924 \\
            {[7, 7, 7]} & 0.12M & {\ul 0.7960} \\
            {[9, 9, 9]} & 0.15M & 0.7951\\
            {[9, 7, 5]}~\textbf{(Proposed)} & 0.12M & \textbf{0.8064} \\
            
            \specialrule{.1em}{.05em}{.05em} 
        \end{tabular}
    }
    \vspace{-1.2\baselineskip}
    
\end{table}

%% file: 4_discussion.tex
\section{Discussion and Conclusion}
\label{sec:discussion}

In this work, we present a novel application of an efficient symmetric rotation-equivariant convolution (\convname) kernel to the task of semantic segmentation.
We integrate this kernel into the standard U-Net framework to learn rotation- and reflection-equivariant segmentation features, while also reducing the model size. 
The effectiveness of our \convabb U-Net is validated on a vessel segmentation task, demonstrating notable improvements across rotated fundus imaging.
Benefiting from the symmetric and rotational equivariance introduced by the \convname kernel, our model can easily detect similar objects that can appear in arbitrary orientations, \eg vessels. This not only allows segmentation models using \convname to better segment target objects but also allows robust performance across rotated data and achieves enhanced performance with a shallower network using fewer trainable parameters. 
Compared with the standard U-Net~\cite{ronneberger2015u} with two downsampling stages, our model has only 25\% of the number of parameters but demonstrates robust performance on vessel segmentation.
Enlarging the test set with small realistic degrees of rotation, we further demonstrate our model can learn subtle rotation-equivariant features, outperforming other equivariant learning methods.
Our \convname U-Net approach surpasses the performance of the SoTA segmentation method~\cite{liu2022full} that utilizes an advanced network structure while using only 1.7\% of the trainable parameters.

The inherent versatility of the \convname kernel, being a fundamental building block for rotation equivariance, can be easily integrated into other existing segmentation frameworks~\cite{schlemper2019attention,zhou2019unet++}.
Given our benchmarking comparisons against the baselines showing a substantial boost in performance across rotation angles, we anticipate a similar improvement upon \convname integration with other deep learning architectures.
This work serves as a proof of concept to demonstrate the feasibility of rotation equivariant semantic segmentation networks.
Addressing rotational equivariance is a critical problem in medical imaging where patient and organ positioning will differ across scans or where image orientation is ambiguous, e.g. histopathology images.
In the future, we aim to integrate \convname kernels into alternative semantic segmentation architectures and rigorously evaluate the efficacy of this approach on other segmentation tasks beyond vessels.

%% file: 5_acknowledgements.tex
\paragraph*{Acknowledgements}
No funding was received to conduct this study. The authors have no relevant financial or non-financial interests to disclose.

\paragraph*{Compliance with Ethical Standards}

This research study was conducted retrospectively using human subject data made available in open access by Joes Staal~\etal~\cite{staal2004ridge}. Ethical approval was not required as confirmed by the license attached with the open-access data.

%% file: 0_main.bbl
\begin{thebibliography}{10}

\bibitem{ronneberger2015u}
Olaf Ronneberger, Philipp Fischer, and Thomas Brox,
\newblock ``U-net: Convolutional networks for biomedical image segmentation,''
\newblock in {\em MICCAI}. Springer, 2015, pp. 234--241.

\bibitem{ji2021learning}
Wei Ji, Yefeng Zheng, et~al.,
\newblock ``Learning calibrated medical image segmentation via multi-rater agreement modeling,''
\newblock in {\em IEEE CVPR}, 2021, pp. 12341--12351.

\bibitem{isensee2021nnu}
Fabian Isensee, Klaus~H Maier-Hein, et~al.,
\newblock ``nnu-net: a self-configuring method for deep learning-based biomedical image segmentation,''
\newblock {\em Nature methods}, vol. 18, no. 2, pp. 203--211, 2021.

\bibitem{zhang2022atlas}
Jiazhen Zhang, John~A Onofrey, et~al.,
\newblock ``Atlas-based semantic segmentation of prostate zones,''
\newblock in {\em MICCAI}. Springer, 2022, pp. 570--579.

\bibitem{dieleman2016exploiting}
Sander Dieleman, Jeffrey De~Fauw, and Koray Kavukcuoglu,
\newblock ``Exploiting cyclic symmetry in convolutional neural networks,''
\newblock in {\em International conference on machine learning}. PMLR, 2016, pp. 1889--1898.

\bibitem{chidester2019rotation}
Benjamin Chidester, Jian Ma, et~al.,
\newblock ``Rotation equivariant and invariant neural networks for microscopy image analysis,''
\newblock {\em Bioinformatics}, vol. 35, no. 14, pp. i530--i537, 2019.

\bibitem{cohen2016group}
Taco Cohen and Max Welling,
\newblock ``Group equivariant convolutional networks,''
\newblock in {\em International conference on machine learning}. PMLR, 2016, pp. 2990--2999.

\bibitem{linmans2018sample}
Jasper Linmans, Max Welling, et~al.,
\newblock ``Sample efficient semantic segmentation using rotation equivariant convolutional networks,''
\newblock {\em arXiv preprint arXiv:1807.00583}, 2018.

\bibitem{winkens2018improved}
Jim Winkens, Max Welling, et~al.,
\newblock ``Improved semantic segmentation for histopathology using rotation equivariant convolutional networks,''
\newblock in {\em MIDL}, 2018.

\bibitem{pang2022beyond}
Shuchao Pang, Zhenmei Yu, et~al.,
\newblock ``Beyond cnns: exploiting further inherent symmetries in medical image segmentation,''
\newblock {\em IEEE transactions on cybernetics}, 2022.

\bibitem{bernander2022rotation}
Karl~Bengtsson Bernander, Ingela Nystr{\"o}m, et~al.,
\newblock ``Rotation-equivariant semantic instance segmentation on biomedical images,''
\newblock in {\em Annual conference on medical image understanding and analysis}. Springer, 2022, pp. 283--297.

\bibitem{worrall2017harmonic}
Daniel~E Worrall, Gabriel~J Brostow, et~al.,
\newblock ``Harmonic networks: Deep translation and rotation equivariance,''
\newblock in {\em IEEE CVPR}, 2017, pp. 5028--5037.

\bibitem{weiler2018learning}
Maurice Weiler, Fred~A Hamprecht, and Martin Storath,
\newblock ``Learning steerable filters for rotation equivariant cnns,''
\newblock in {\em IEEE CVPR}, 2018, pp. 849--858.

\bibitem{weiler2019general}
Maurice Weiler and Gabriele Cesa,
\newblock ``General e (2)-equivariant steerable cnns,''
\newblock {\em Advances in neural information processing systems}, vol. 32, 2019.

\bibitem{cesa2022a}
Gabriele Cesa, Leon Lang, and Maurice Weiler,
\newblock ``A program to build {E(N)}-equivariant steerable {CNN}s,''
\newblock in {\em ICLR}, 2022.

\bibitem{dudar2019use}
Viacheslav Dudar and Vladimir Semenov,
\newblock ``Use of symmetric kernels for convolutional neural networks,''
\newblock in {\em ICDSIAI}. Springer, 2019, pp. 3--10.

\bibitem{fuhl2021rotated}
Wolfgang Fuhl and Enkelejda Kasneci,
\newblock ``Rotated ring, radial and depth wise separable radial convolutions,''
\newblock in {\em IJCNN}. IEEE, 2021, pp. 1--8.

\bibitem{du2025sreconv}
Yuexi Du, John~A Onofrey, et~al.,
\newblock ``Sre-conv: Symmetric rotation equivariant convolution for biomedical image classification,''
\newblock {\em arXiv preprint arXiv:2501.09753}, 2025.

\bibitem{staal2004ridge}
Joes Staal, Bram Van~Ginneken, et~al.,
\newblock ``Ridge-based vessel segmentation in color images of the retina,''
\newblock {\em IEEE transactions on medical imaging}, vol. 23, no. 4, pp. 501--509, 2004.

\bibitem{liu2022full}
Wentao Liu, Feng Gao, et~al.,
\newblock ``Full-resolution network and dual-threshold iteration for retinal vessel and coronary angiograph segmentation,''
\newblock {\em IEEE journal of biomedical and health informatics}, vol. 26, no. 9, pp. 4623--4634, 2022.

\bibitem{frangi1998multiscale}
Alejandro~F Frangi, Max~A Viergever, et~al.,
\newblock ``Multiscale vessel enhancement filtering,''
\newblock in {\em MICCAI}. Springer, 1998, pp. 130--137.

\bibitem{schlemper2019attention}
Jo~Schlemper, Daniel Rueckert, et~al.,
\newblock ``Attention gated networks: Learning to leverage salient regions in medical images,''
\newblock {\em Medical image analysis}, vol. 53, pp. 197--207, 2019.

\bibitem{zhou2019unet++}
Zongwei Zhou, Jianming Liang, et~al.,
\newblock ``Unet++: Redesigning skip connections to exploit multiscale features in image segmentation,''
\newblock {\em IEEE transactions on medical imaging}, vol. 39, no. 6, pp. 1856--1867, 2019.

\end{thebibliography}
